\documentclass[aps,pra,twocolumn,groupedaddress,showpacs]{revtex4}

\usepackage{graphicx}
\usepackage{dcolumn}
\usepackage{bm}
\usepackage{verbatim}
\usepackage{amsmath,amssymb}
\usepackage{relsize,exscale}
\usepackage{color}
\usepackage{hyperref}
\begin{document}

\title{Ehrenfest principle and unitary dynamics of quantum-classical systems
with general potential interaction}

\author{M. Radonji\'c}
\author{D. B. Popovi\' c}
\author{S. Prvanovi\'c}
\author{N. Buri\'c}
\email[]{buric@ipb.ac.rs}
\affiliation{Institute of Physics, University of Belgrade, Pregrevica 118, 11080 Belgrade, Serbia}

\begin{abstract}

Representation of classical dynamics by unitary transformations has been used to develop unified
description of hybrid classical-quantum systems with particular type of interaction, and to formulate
abstract systems interpolating between classical and quantum ones. We solved the problem of unitary
description of two interpolating systems with general potential interaction. The general solution is
used to show that with arbitrary potential interaction between the two interpolating systems the
evolution of the so called unobservable variables is decoupled from that of the observable ones if
and only if the interpolation parameters in the two interpolating systems are equal.

\end{abstract}


\maketitle

\section{Introduction}

Koopman-von Neumann (KvN) \cite{KN} unitary description of the Liouville equation of classical
Hamiltonian dynamical systems was utilized for modeling hybrid quantum-classical systems for the
first time by Sherry and Sudarshan \cite{Sud}. They analyzed particular types of interaction between
the classical and the quantum parts, and ad hoc prescriptions for definitions of the corresponding
Hilbert space operators. It was shown that the pre-measurement process can be modeled as an
interaction between a classical apparatus and a quantum system within the unitary framework. Sherry
and Sudarshan also analyzed the so called integrity conditions which ought to be satisfied in order
that classical variables remain classical during the hybrid unitary evolution in the Heisenberg form.
Peres and Terno \cite{Peres} analyzed consistency of the Koopman-von Neumann-Sudarshan (KNS) hybrid
dynamics with the quantum-quantum and the classical-classical limits for the case of linear
interaction between harmonic oscillators. Some aspects of the KNS formalism for hybrid system with
specific interaction have also been studied in \cite{spanci}. The authors investigated the role of
unphysical variables which are called unobservables because they do not influence the evolution of
the physical observables of the quantum or the classical part if there is no quantum-classical
interaction. It was observed, using particular examples of quantum-classical interaction and specific
forms of its Hilbert space description, that the evolution of the unobservable and observable
variables become coupled.

More recently, KvN formalism  and Ehrenfest principle  were used to propose a family of abstract unitary systems interpolating
between classical system and its quantized counterpart \cite{oper}. The problem of hybrid dynamics
was not analyzed using the interpolating systems. Our goal is to study the same type of questions,
but for the most general potential interaction between the classical and the quantum systems. In
fact, we shall obtain unitary dynamical equations for two interpolating abstract systems (IAS) with
general potential interaction, and use this to show that generally the evolution of the unphysical
variables is decoupled from that of the physical ones if and only if the interpolation parameters in
the two IAS are equal. In particular, unitary dynamics of hybrid systems with potential interaction
in general couples the dynamics of the two types of variables. However, there is one special case in
the family of general solutions such that the corresponding quantum-classical potential interaction
does not couple the physical and the unphysical variables, and implies other properties consistent
with this fact.

\section{Interpolating abstract systems and hybrid models}

Dynamical equations for averages of the basic observables of a classical system and that of its
quantized counterpart can be mathematically interpolated by an abstract system that depends on
suitable parameter. The first step to achieve this is to rewrite the dynamics of classical and
quantum averages using the same mathematical framework. This can be done by rewriting the classical
dynamics as a unitary evolution on a suitable Hilbert space, or by rewriting the unitary
Schr\"odinger equation as a (linear) Hamiltonian system on a symplectic manifold. We shall treat here
the unitary approach with general potential interaction.

Consider an abstract dynamical system with the basic variables $x_j,p_j,\chi_j,\pi_j$ (hereafter
$j=1,2$). Properties of the system, expressed through appropriate algebraic relations between the
basic variables, are supposed to depend on parameters $a_j$. The basic variables satisfy commutation
relations
\begin{equation}\label{VCs}
[x_j,p_j]=i\hbar\;\! a_j,\quad [x_j,\pi_j]=[\chi_j,p_j]=i\hbar,
\end{equation}
with all other commutators being zero. Let us suppose that the algebra (\ref{VCs}) is represented by
operators acting on a Hilbert space ${\cal H}$. Assume that the dynamical variables $x_j,p_j$
are measurable and that their averages in a state $|\psi\rangle\in {\cal H}$ are computed as
\begin{equation}\label{aver}
\langle x_j\rangle_\psi=\langle \psi|\hat x_j|\psi\rangle,\quad
\langle p_j\rangle_\psi=\langle \psi|\hat p_j|\psi\rangle.
\end{equation}
Suppose that the dynamics of these averages is given by the Ehrenfest principle
\begin{subequations}\label{Et_dec}
\begin{align}
\frac{d}{dt}\langle\psi(t)|\hat x_j|\psi(t)\rangle&=\langle\psi(t)|\frac{\hat p_j}{m_j}|\psi(t)
\rangle, \\
\frac{d}{dt}\langle\psi(t)|\hat p_j|\psi(t)\rangle&=\langle\psi(t)|-V_j'(\hat x_j)|\psi(t)\rangle,
\end{align}
\end{subequations}
and that the state evolution is unitary $i\hbar |\dot\psi\rangle=\hat H_{I\!AS}|\psi\rangle$. The
corresponding evolution equations for the dynamical variables in the Heisenberg form are
$i\hbar\,d\hat x_j/dt=[\hat x_j, \hat H_{I\!AS}]$ and analogously for $\hat p_j,\hat\chi_j,
\hat\pi_j$. The operator $\hat H_{I\!AS}$ is the evolution generator and might depend on all
dynamical variables $\hat H_{I\!AS}=H_{I\!AS}(\hat x_j,\hat p_j,\hat\chi_j,\hat\pi_j)$. It is not
necessarily interpreted as the physical energy. It should be remarked that the relations (\ref{aver})
and (\ref{Et_dec}) are treated as axioms in the general abstract formulation \cite{oper}, expressing
the conservative nature of the dynamics. \mbox{Following} the approach of \cite{oper}, one can obtain
the class of evolution generators yielding (\ref{Et_dec})
\begin{align}\label{H_abs}
\hat H_{I\!AS} &=\sum_{j=1,2} \frac{1}{a_j}\bigg(\frac{\hat p_j^2}{2m_j}+V_j^{}(\hat x_j)\bigg)
\nonumber \\  &+ F_j(\hat x_j-a_j\:\!\hat\chi_j,\hat p_j-a_j\:\!\hat\pi_j),
\end{align}
where $F_j$ are arbitrary functions of the indicated arguments. Observe that, consistent with
(\ref{Et_dec}), there are no terms coupling observables with different subscripts, so that the
abstract system (\ref{H_abs}) can be interpreted as a compound system with two noninteracting
components.

Explicit representation of the operator $\hat H_{I\!AS}$ depends on the representation space ${\cal
H}$, and is not important in our analysis. Nevertheless, it should be remarked that the Hilbert space
${\cal H}$ is determined as a space of an irreducible representation of the algebra (\ref{VCs}), and
is the same space for any value of the parameters $a_j$. In particular, it is seen that in the case
we want to represent two quantum systems, the Hilbert space needed to accommodate (\ref{VCs}) with
$a_1=a_2=1$ is larger than the space $L_2(x_1)\otimes L_2(x_2)\equiv L_2(x_1,x_2)$ which is relevant
in the standard quantum mechanics without the additional variables $\chi_j,\pi_j$. It can be shown
that one irreducible representation of the algebra is provided with the Hilbert space of operators on
$L_2(x_1,x_2)$ \cite{Mukunda}. Thus, the vectors from ${\cal H}$ can be considered as density
matrices or mixed states of the quantum-quantum system \cite{oper}. Similarly, if the abstract
systems represent two classical systems, i.e.\ when $a_1=a_2=0$ so that $\hat x_j,\hat p_j$ all
commute, the interpretation of the state $|\psi\rangle$ is that of the amplitude of a probability
density $\rho(x_1,x_2,p_1,p_2)=|\langle x_1,x_2,p_1,p_2|\psi\rangle|^2$ on the corresponding phase
space ${\cal M}(x_1,x_2,p_1,p_2)$ \cite{oper}. The scalar product in (\ref{aver}) coincides with
the ensemble average $\int_{\cal M} \rho\,x_j\,dM$ or $\int_{\cal M} \rho\,p_j\,dM$. Observe that
the classical Hilbert space can be partitioned into equivalence classes $|\psi\rangle\sim
e^{i\phi}|\psi\rangle$, where each class corresponds to a single density $\rho$. The evolution
equations preserve the equivalence classes because there is no interaction \cite{spanci}.

Convenient choices of the arbitrary functions $F_j$ can reproduce the evolution equations for
non-interacting classical-classical (C-C) $(a_1=a_2=0)$, quantum-quantum \mbox{(Q-Q)} $(a_1=a_2=1)$
and classical-quantum systems (C-Q) $(a_1=0$, $a_2=1)$. The relevant choice of functions $F_j$ and
the corresponding equations can be obtained as the special case from the general equations, that will
be given later, with interaction set to zero.

For arbitrary $a_1,a_2\neq 0,1$ the dynamical equations describe the evolution of an abstract
system interpolating between the quantum and the classical systems (hence the notation $\hat
H_{I\!AS}$). Because there is no interaction between the two systems, the evolution of $\hat x_j,
\hat p_j$ is also independent of $\hat\chi_j,\hat\pi_j$. The system has 2+2 degrees of freedom, and
each of the degrees of freedom evolves independently of the others. If the abstract system
(\ref{H_abs}) is meant to represents two quantum or two classical systems, the variables $\hat
x_j,\hat p_j$ are interpreted as physical observables of coordinates and momenta. The variables
$\hat\chi_j,\hat\pi_j$, similarly as $\hat H_{I\!AS}$, do not represent physical observables. They
are dynamically separated from the physical observables and appear because the family of systems
(\ref{H_abs}) must interpolate between the classical and the quantum dynamics \cite{oper}.


\section{IAS with general potential interaction}

Potential interaction between two quantum systems or between two classical systems appears in the
equations of motion in the form of gradients of the corresponding scalar potential. In the extended
Hilbert space formalism, which is required for the formulation of the IAS, such potential Q-Q or C-C
interaction can be represented by an operator expression in terms of all variables with the role of
coordinates $\hat W=W(\hat x_1,\hat x_2,\hat\chi_1,\hat\chi_2)$. We assume that in the dynamical
equations for the corresponding momenta $\hat W$ should appear as gradient with respect to the
corresponding coordinate.

We shall now consider dynamics of two abstract systems with arbitrary values of $a_1,a_2$ and with
an arbitrary potential interaction between them. Like in the Q-Q and C-C cases, we demand that the
following relations hold
\begin{subequations}\label{Et_int}
\begin{align}
\frac{d}{dt}\langle\Psi(t)|\hat x_j|\Psi(t)\rangle&=\langle\Psi(t)|
\frac{\hat p_j}{m_j}|\Psi(t)\rangle,\\
\frac{d}{dt}\langle\Psi(t)|\hat p_j|\Psi(t)\rangle&=\langle\Psi(t)|-V_j'(\hat x_j)
-\frac{\partial\hat W}{\partial\hat x_j}|\Psi(t)\rangle.
\end{align}
\end{subequations}
Notice that the potential interaction can be completely general. Particular examples of
interaction which do not necessarily satisfy (\ref{Et_int}) have been assumed in somewhat ad hoc
manner and studied in \cite{Sud,Peres,spanci}. Our goal is to determine the unitary evolution
generator $\hat H_{I\!AS}=H_{I\!AS}(\hat x_1,\hat p_1,\hat x_2,\hat p_2,\hat\chi_1,\hat\pi_1,
\hat\chi_2,\hat\pi_2)$ such that $i\hbar\:\!|d\Psi(t)/dt\rangle=\hat H_{I\!AS}|\Psi(t)\rangle$ holds.
The unitary evolution and (\ref{Et_int}) give the following relations
\begin{align}\label{CRs}
\frac{1}{i\hbar}[\hat x_j,\hat H_{I\!AS}]=\frac{\hat p_j}{m_j},\quad
-\frac{1}{i\hbar}[\hat p_j,\hat H_{I\!AS}]=V_j'(\hat x_j)+\frac{\partial\hat W}{\partial\hat x_j},
\end{align}
and the related system of partial differential equations (PDEs) for the function $ H_{I\!AS}$,
\begin{subequations}\label{PDEs}
\begin{align}
a_j\frac{\partial H_{I\!AS}}{\partial p_j}+\frac{\partial H_{I\!AS}}{\partial\pi_j}
&=\frac{p_j}{m_j},\\
a_j\frac{\partial H_{I\!AS}}{\partial x_j}+\frac{\partial H_{I\!AS}}{\partial\chi_j}
&=V_j'(x_j)+\frac{\partial W}{\partial x_j}.
\end{align}
\end{subequations}
The commutation relations (\ref{CRs}), i.e.\ the PDEs (\ref{PDEs}), are not consistent for
arbitrary choice of the interaction potential $\hat W$. Jacobi identity $[\hat H_{I\!AS},[\hat
p_1,\hat p_2]]+[\hat p_1,[\hat p_2,\hat H_{I\!AS}]]+[\hat p_2,[\hat H_{I\!AS},\hat p_1]]=0$ and the
commutation relation $[\hat p_1,\hat p_2]=0$ imply that $[\hat p_1,[\hat p_2,\hat H_{I\!AS}]]=[\hat
p_2,[\hat p_1,\hat H_{I\!AS}]]$, so that
\begin{align}
\left[a_1\frac{\partial}{\partial x_1}+\frac{\partial}{\partial\chi_1},
a_2\frac{\partial}{\partial x_2}+\frac{\partial}{\partial\chi_2}\right]{ H_{I\!AS}}=0
\end{align}
must be satisfied. Invoking the second relation of (\ref{PDEs}), we get the consistency requirement
\begin{equation}\label{Req}
\left(a_1\frac{\partial}{\partial x_1}+\frac{\partial}{\partial\chi_1}\right)
\frac{\partial W}{\partial x_2} -
\left(a_2\frac{\partial}{\partial x_2}+\frac{\partial}{\partial\chi_2}\right)
\frac{\partial W}{\partial x_1}=0.
\end{equation}
The general solution of (\ref{Req}) is
\begin{equation}\label{GenSol}
W = \int_{-\infty}^\infty{\cal W}(x_1+(\alpha-a_1)\chi_1,\,x_2+(\alpha-a_2)\chi_2,\,\alpha)\,d\alpha,
\end{equation}
where $\cal W$ is an arbitrary function such that the previous integral is defined. Note that when
$a_1\neq a_2$, i.e.\ when the systems are of different type, the interaction potential $\hat W$ will
depend on at least one of the unobservables $\hat\chi_1,\hat\chi_2$. This conclusion remains valid
in the particular case of hybrid classical-quantum system, where $a_1=0$ corresponds to the classical
part and $a_2=1$ is related to the quantum part. Let us stress that this fact is proved here for
quite general potential interaction and not just observed for some special choices of the
interaction \cite{Peres,spanci}.

Consider a particular choice of ${\cal W}\propto\delta(\alpha-a)$ yielding the interaction potential
$\hat W=W(\hat x_1+(a-a_1)\hat\chi_1,\,\hat x_2+(a-a_2)\hat\chi_2)$. The related solution of the PDEs
(\ref{PDEs}) gives
\begin{align}\label{Ha1a2}
\hat H_{I\!AS}&=\sum_{j=1,2}\frac{1}{a_j}\bigg(\frac{\hat p_j^2}{2m_j}+V^{}_j(\hat x_j)\bigg)
\nonumber\\
&+\frac{1}{a}\,W(\hat x_1+(a-a_1)\hat\chi_1,\,\hat x_2+(a-a_2)\hat\chi_2)\nonumber\\
&+F(\hat x_1-a_1\:\!\hat\chi_1,\hat p_1-a_1\:\!\hat\pi_1,\hat x_2-a_2\:\!\hat\chi_2,\hat
p_2-a_2\:\!\hat\pi_2),
\end{align}
where $F$ is arbitrary real-valued smooth function that commutes with the observables
$O(\hat x_1,\hat p_1,\hat x_2,\hat p_2)$. Let us observe that
when the two systems are of the same type, the unobservables do not influence the evolution of the
physical observables for the choice $a_1=a_2=a$. The result (\ref{Ha1a2}) can be extended, although
with some care, to the limit $a\to 0$, which will turn out to be interesting for the hybrid Q-C
system. Namely, one can take a part of the function $F$ to be of the suitable form
$-\frac{1}{a}\,W(\hat x_1-a_1\hat\chi_1,\,\hat x_2-a_2\hat\chi_2)$ that yields in the $a\to 0$ limit,
\begin{align}
\hat H_{I\!AS}&=\sum_{j=1,2}\frac{1}{a_j}\bigg(\frac{\hat p_j^2}{2m_j}+V^{}_j(\hat x_j)\bigg)
\nonumber\\
&+\partial_1 W(\hat x_1-a_1\hat\chi_1,\,\hat x_2-a_2\hat\chi_2)\,\hat\chi_1\nonumber\\
&+\partial_2 W(\hat x_1-a_1\hat\chi_1,\,\hat x_2-a_2\hat\chi_2)\,\hat\chi_2\nonumber\\
&+F(\hat x_1-a_1\:\!\hat\chi_1,\hat p_1-a_1\:\!\hat\pi_1,\hat x_2-a_2\:\!\hat\chi_2,\hat
p_2-a_2\:\!\hat\pi_2),
\end{align}
where $\partial_j W$ denotes partial derivative of the potential with respect to the $j$-th
argument.

The limit of (\ref{Ha1a2}) when $a_j\to 0$ can also be obtained by choosing a part of the function
$F$ in the form $-\frac{1}{a_j}\big(\frac{(\hat p_j-a_j\hat\pi_j)^2} {2m_j}+V^{}_j(\hat
x_j-a_j\hat\chi_j)\big)$, as in \cite{oper}. In particular, this yields the
Hamiltonian, as the dynamics generator, of a hybrid classical-quantum system $(a_1\to 0$, $a_2=1)$
\begin{align}\label{Hyb}
\hat{H}_{\rm hyb}&=\frac{\hat p^{}_1}{m_1}\,\hat\pi_1+V_1'(\hat x_1)\,\hat\chi_1+
\frac{\hat p_2^2}{2m_2}+V^{}_2(\hat x_2)\nonumber\\
&+\frac{1}{a}\,W(\hat x_1+a\hat\chi_1,\,\hat x_2+(a-1)\hat\chi_2)\nonumber\\
&+F(\hat x_1,\hat p_1,\hat x_2-\:\!\hat\chi_2,\hat p_2-\:\!\hat\pi_2),
\end{align}
where the first four terms describe non-interacting hybrid system. As already mentioned, the
interaction potential depends on at least one of the unobservables $\hat\chi_1,\hat\chi_2$. The
appearance of the unphysical variables in the Hamiltonian is not a problem {\it per se}, because the
Hamiltonian is anyway interpreted as the dynamics generator and not as the physical energy.
Additionally, in the purely C-C case $(a_1=a_2=a\to 0)$ one gets
\begin{align}
\hat{H}_{c\text{-}c}&=\frac{\hat p^{}_1}{m_1}\,\hat\pi_1+V_1'(\hat x_1)\,\hat\chi_1+
\frac{\hat p^{}_2}{m_2}\,\hat\pi_2+V_2'(\hat x_2)\,\hat\chi_2\nonumber\\
&+\partial_1 W(\hat x_1,\hat x_2)\,\hat\chi_1+\partial_2 W(\hat x_1,\hat x_2)\,\hat\chi_2,
\end{align}
with the unobservables being present, but not within the arguments of the interaction potential.
However, in C-C case the unphysical variables do not appear in the evolution equations of the
physical observables. We may remark in passing that $\hat W$ is not interpreted as the potential
energy of the hybrid, but as a term in the generator of dynamics corresponding to the potential
interaction. However, the crucial property of hybrid Q-C systems is that the equations of motion for
the physical and unphysical variables become coupled. Those equations are easily obtained from the
generator (\ref{Hyb}). Thus, we have shown that the dynamical equations couple physical and
unphysical variables in the case of potential Q-C interaction in general, that is with the
Hamiltonian of the general hybrid form (\ref{Hyb}).

A very special case of (\ref{Hyb}) is obtained in the limit $a\to 0$ with the appropriate choice of
the function $F$ yielding the Hamiltonian
\begin{align}
\hat{ H}_{\rm hyb}&=\frac{\hat p^{}_1}{m_1}\,\hat\pi_1+V_1'(\hat x_1)\,\hat\chi_1+
\frac{\hat p_2^2}{2m_2}+V^{}_2(\hat x_2)\nonumber\\
&+\partial_1 W(\hat x_1,\,\hat x_2-\hat\chi_2)\,\hat\chi_1
+\partial_2 W(\hat x_1,\,\hat x_2-\hat\chi_2)\,\hat\chi_2,
\end{align}
with the corresponding equations of motion of the variables
\begin{subequations}
\begin{align}
&\frac{d\hat x_j}{dt} = \frac{\hat p^{}_j}{m_j},\\
&\frac{d\hat p_j}{dt} = -V_j'(\hat x_j)-\partial_j W(\hat x_1,\,\hat x_2-\hat\chi_2),\\
&\frac{d}{dt}(\hat x_2-\hat\chi_2)=0.
\end{align}
\end{subequations}
This solution describes the situation when the evolution of the classical
system depends on the quantum system only through a constant of motion $\hat x_2-\hat\chi_2$.
In this very special case of the general hybrid solution, the classical variables see only a quite
coarse-grained effect of the quantum evolution. On the other hand, the dynamics of the quantum
sector is influenced by the details of the dynamics of the classical physical variables $\hat x_1,
\hat p_1$. In addition, this is the only case of the potential Q-C interaction which satisfies
the integrity principle of Sudarshan \cite{Sud}. Namely, the terms $\partial_j W(\hat x_1,\,\hat
x_2-\hat\chi_2)$ in this form of the Hamiltonian commute with the momenta $\hat p_1,\hat p_2$,
which assures commutation of the classical variables at different times.

\section{Summary}

We have studied the type of theory of hybrid quantum-classical systems where the evolution is
described by unitary transformations on an appropriate Hilbert space. The fact that both
classical and quantum mechanics can be formulated on the same Hilbert space makes it possible to
introduce a parameter dependent family of abstract systems interpolating between a classical system
and its quantized counterpart \cite{oper}. The variables involved in the formulation of the abstract
interpolating model can be divided into two groups, one with the standard physical interpretation and
one with no physical interpretation. In the limits of the classical or the quantum system the two
groups of variables are dynamically separated. We have studied two such abstract interpolating
systems with quite arbitrary potential interaction between them. General solution for the problem of
constructing dynamical equations for such a pair of systems is provided for the first time. It is
shown that, with the most general type of potential interaction, the dynamics of the two groups of
variables is separated if and only if the two abstract interpolating systems have the same value of
the interpolation parameter. On the other hand, if the interpolation parameters of the two system are
different, the two groups of variables dynamically influence each other. The variables which can be
considered as unphysical and cannot be observed in the purely quantum or in the purely classical
case, do have an observable effect in the hybrid quantum-classical system. Our results demonstrate
this fact for arbitrary potential interaction, in line with the previous special cases
\cite{Peres,spanci}. Analogous conclusions are obtained in the symplectic approach to the
conservative hybrid dynamics \cite{Elze,usPRA}, and the analogy is worth further investigation. We
have also analyzed the particular case of the general solution corresponding to the situation when
the classical part is influenced by the quantum part only through a particular combination of the
variables from the quantum system that remains constant during the evolution. This, rather special
case, is the only possible dynamics of the hybrid system within the framework of unitary dynamics
with potential interaction, when the physical and the unphysical variables can be considered as
decoupled, and also when the \mbox{Sudarshan} integrity condition of the classical system is
satisfied.

\begin{acknowledgments}
We acknowledge support of the Ministry of Education and Science of the Republic of Serbia, contracts
No.\ 171006, 171017, 171020, 171038 and 45016 and COST (Action MP1006).
\end{acknowledgments}

\end{document}